# Uniqueness of Herndon's Georeactor: Energy Source and Production Mechanism for Earth's Magnetic Field

by

J. Marvin Herndon
Transdyne Corporation
San Diego, CA 92131 USA
Herndon@NuclearPlanet.com

**Abstract:** Herndon's georeactor at the center of Earth is immune to meltdown, which is not the case for recently published copy-cat georeactors, which would necessarily be subject to "hot" nuclear fuel, prevailing high-temperature environments, and high confining pressures. Herndon's georeactor uniquely is expected to be self-regulating through establishing a balance between heat-production and actinide settling-out. The seventy year old idea of convection in the Earth's fluid core is refuted because thermal expansion cannot overcome the 23% higher density at the core's bottom than at its top. The dimensionless Rayleigh Number is an inappropriate indicator of convection in the Earth's core and mantle as a consequence of the assumptions under which it was derived. Implications bearing on the origin of the geomagnetic field, the physical impossibility of mantle convection, and the concomitant refutation of plate tectonics theory are briefly described.

In 1993 and 1994, Herndon [1, 2] published the concept and applied Fermi's nuclear reactor theory [3] to demonstrate the feasibility of a naturally occurring nuclear fission at the center of the Earth, now called the georeactor, as the energy source for the geomagnetic field. In 1996, Herndon [4] disclosed the sub-structure of the inner core, describing the two-component structure of the georeactor as consisting of an actinide sub-core, surrounded by a sub-shell composed of the products of nuclear fission and radioactive decay, all surrounded by the inner core. He also noted the possibility that the sub-shell might be a liquid or slurry.

In 2001, Hollenbach and Herndon [5] published the first georeactor numerical simulation conducted using Oak Ridge National Laboratory's SCALE software, which has been validated with nuclear reactor operating data combined with analyses of spent fuel rods [6]. The numerical simulations showed that georeactor-produced $^3$He and $^4$He would have the same range of compositions as helium measured in oceanic basalts [7]. Subsequently, Herndon [8] published more precise numerical simulation data, examples of which are shown in Figure 1 from [9]. The marked, progressive increase in $^3$He/$^4$He ratios over time occurs as a consequence of diminished $^4$He production from radioactive decay as uranium is consumed. Herndon suggested that the high $^3$He/$^4$He ratios observed in Hawaiian and Icelandic lavas [10] portend the demise of the georeactor, although the time scale is uncertain [11].

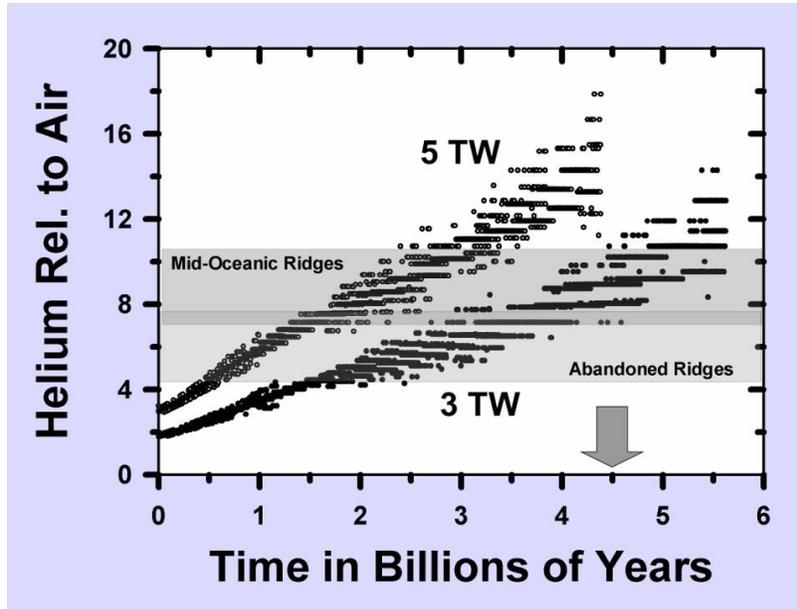

**Figure 1.** Georeactor numerical simulation $^3$He/$^4$He ratios, relative to those in air, over the age of the Earth, calculated at power levels of 3 and 5 terawatts (TW). Arrow indicates present age of Earth. Range of measured oceanic basalt helium ratios at 95% confidence intervals are shown for two major oceanic provinces. Note that helium ratios as high as 37 relative to air are observed in some samples of Icelandic and Hawaiian lavas.

As individuals started to appreciate the importance of the georeactor concept, "copy-cat" georeactors began to be published in the scientific literature. These all possess the common feature of supposedly occurring at places other than at the center of the Earth, specifically, at the core-mantle boundary [12] and at the surface of the inner core [13, 14]. All copy-cat georeactors are absence one consideration, which is their common Achilles heel – the potential for meltdown.

Figure 2 shows the $^{235}$U/$^{238}$U ratio of Herndon's georeactor over the age of Earth from numerical simulation calculations [8]. For reference, the $^{235}$U/$^{238}$U ratio is also shown for (*i*) the Oklo natural nuclear reactor in the Republic of Gabon at time of its operation, (*ii*) the Chernobyl nuclear reactor which suffered meltdown and (*iii*) natural uranium at present time. Note the high georeactor $^{235}U/^{238}U$ ratios, especially during the first 1.5 billion years of Earth's existence. Even at present, that ratio is still "hot" from a nuclear fission standpoint as compared, for example, to Chernobyl fuel. Such high $^{235}U/^{238}U$ ratios strongly enhance copy-cat georeactor meltdown, which would lead to the dense uranium progressing downward to Earth's center, if not being diluted by the molten underlying matter until criticality ceases.

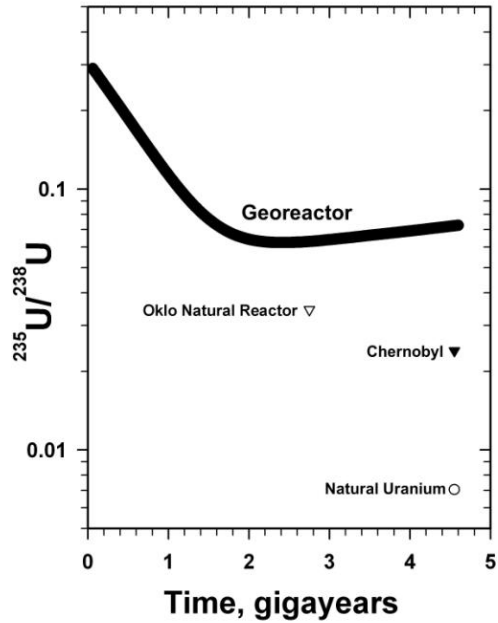

**Figure 2.** Georeactor $^{235}U/^{238}U$ ratios over the age of the Earth. Reference points are shown for the Oklo natural reactor, Chernobyl reactor, and natural uranium at present.

Copy-cat georeactor meltdown to the location of Herndon's georeactor appears inevitable as no nuclear fission control mechanism for these hypothetical reactors is known. Unlike, for example, the Earth-surface natural reactor at Oklo which did not experience meltdown as it began to function with a significantly lower $^{235}U/^{238}U$ ratio in a low-temperature environment and functioned primarily as a thermal neutron reactor, moderated by and pulse-controlled by easily volatilized groundwater [15].

Copy-cat georeactor meltdown is also enhanced by the high-temperatures that prevail adjacent to Earth's molten iron alloy core. Moreover, the pressures that prevail at the core-mantle-boundary (140 GPa) and atop the inner core (330 GPa) tend to oppose copy-cat reactor disassembly, further enhancing their potential for meltdown.

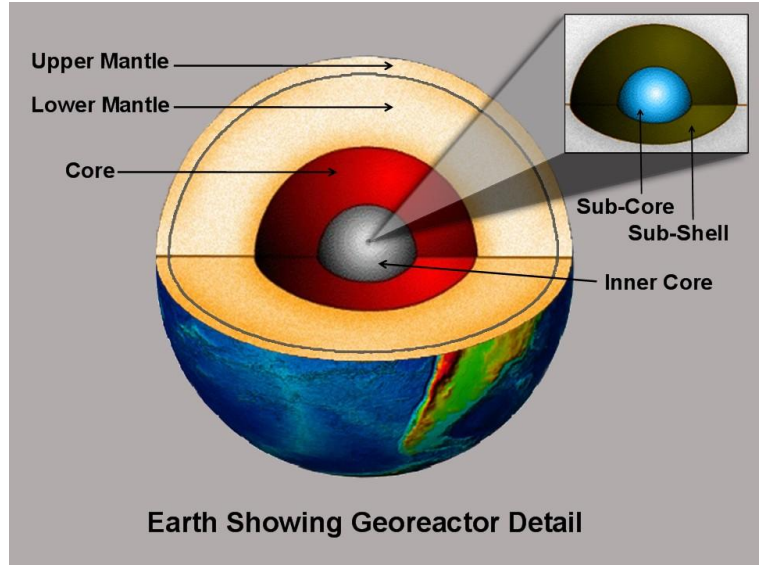

**Figure 3**. Schematic view of Earth's interior showing Herndon's georeactor.

Figure 3 from [9] is a schematic representation of the parts of the interior of Earth, showing Herndon's georeactor. Note that the actinide fuel lies at the gravitational center, the ultimate meltdown-endpoint. Herndon [16] recently published evidence that the neutron-absorbing sub-shell, surrounding the actinide sub-core, is a liquid or slurry. He also pointed out that, because the surrounding inner core is a massive, thermally-conducting heat sink, surrounded by the core, an even more massive thermally-conducting heat sink, long-term-stable convection is to be expected in the georeactor sub-shell. Moreover, as suggested here, sub-shell convection provides a mechanism for enhancing georeactor stability.

In the micro-gravity environment at the center of Earth, georeactor heat production that is too energetic would be expected to cause actinide sub-core disassembly, mixing actinide elements with neutron-absorbers of the sub-shell, quenching the nuclear fission chain reaction. But as the denser actinide elements begin to settle out of the mix, the chain reaction would re-start, ultimately establishing a balance, an equilibrium between heat-production and actinide settling-out, a self-regulating control mechanism.

Seventy years ago, Elsasser [17] published his idea that the geomagnetic field is produced by convective motions in the Earth's fluid, electrically-conducting core, interacting with Coriolis forces produced by planetary rotation, creating a dynamo mechanism, a magnetic amplifier. From studies of rock-magnetism [18], it is known that the geomagnetic field has existed for at least 3.5 billion years. The geomagnetic field has been remarkably stable for long periods of time, including intervals of more than 40 million years without reversals.

Chandrasekhar [19] described convection in the following way: "The simplest example of thermally induced convection arises when a horizontal layer of fluid is heated from below and an adverse temperature gradient is maintained. The adjective 'adverse' is used to qualify the prevailing temperature gradient, since, on account of thermal expansion, the fluid at the bottom becomes lighter than the fluid at the top; and this is a top-heavy arrangement which is potentially unstable. Under these circumstances the fluid will try to redistribute itself to redress this weakness in its arrangement. This is how thermal convection originates: It represents the efforts of the fluid to restore to itself some degree of stability." Understanding the clarity of Chandrasekhar's explanation has led to two reasons that convection in the Earth's fluid core is physically impossible.

Recently, Herndon [16] pointed out that the Earth's fluid core is wrapped in an insulating blanket, a rock shell, the mantle, that is about 2900 km thick, and which has a considerably lower heat capacity, lower thermal conductivity, and higher viscosity than the fluid core [16]. Heat brought to the top of the core cannot be efficiently removed, so maintaining a significant difference in temperature between top and bottom of the core for extended periods of time, a requisite condition for long-term convection, is not possible.

Frequent reference is (wrongly) made to the parameters of the Earth's core yielding a high Rayleigh Number supposedly indicative of vigorous convection. In 1916, Lord Rayleigh [20] applied the Boussinesq [21] approximation to Eulerian equations of motion to derive that dimensionless number to quantify the onset of instability in a thin, horizontal layer of fluid heated from beneath. The underlying assumptions, however, are inconsistent with the physical parameters of the Earth's core, viz.; Earth's core being "incompressible", density being "constant" except as modified by thermal expansion, and pressure being "unimportant" (quotes from Lord Rayleigh [20]).

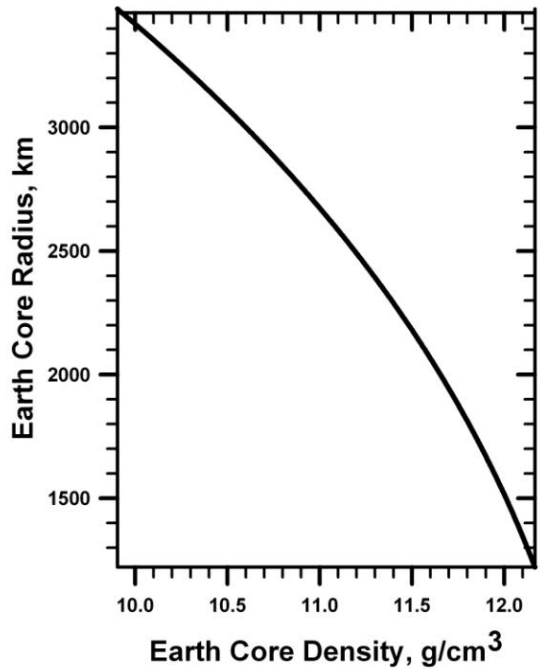

**Figure 4.** Density as a function of radius in the Earth's core [22].

As shown in Figure 4, because of the over-burden weight, the Earth's core is about 23% more dense at the bottom than at the top [22]. Thermal expansion at the bottom cannot overcome such a great difference in density, meaning the Earth's core cannot become top-heavy, and thus cannot engage in convection. The implication is quite clear: Either the geomagnetic field is generated by a process other than the convection-driven dynamo-mechanism, or there exists another fluid region within the deep-interior of Earth which can sustain convection for extended periods of time. Herndon [16] has provided the reasonable basis to expect long-term stable convection in the georeactor sub-shell, and proposed that the geomagnetic field is generated therein by the convection-driven dynamo mechanism.

There are fundamental differences in convection-driven dynamo action in the georeactor sub-shell and in the Earth's core, as has long been believed. First, the georeactor sub-shell contains a substantial quantity of continuously-supplied, neutron-rich, radioactive fission products that beta decay, producing electrons which can generate magnetic seed-fields for amplification. Second, the dimensions, mass, and inertia are orders of magnitude less than those of the Earth's core, meaning that changes in the geomagnetic field, including reversals and excursions [23], can take place on much shorter time-scales than previously thought. Similarly, external effects may assume greater importance. For example, one may speculate that super-intense bursts of solar wind might induce electrical currents and consequently ohmic heating in the georeactor sub-shell, perhaps destabilizing convection and leading to magnetic reversals.

Herndon's georeactor at the center of Earth is immune to meltdown, which is not the case for copy-cat georeactors, which would necessarily be subject to "hot" nuclear fuel, prevailing high-temperature environments, and high confining pressures. Herndon's georeactor uniquely is expected to be self-regulating through establishing a balance between heat-production and actinide settling-out. The seventy year old idea of convection in the Earth's fluid core is refuted because thermal expansion cannot overcome the 23% higher density at the core's bottom than at its top.

The basis presented here for the absence of Earth core convection can be generalized to the Earth's mantle. Since the 1960's, convection has been widely assumed to occur within the Earth's mantle and has been incorporated as an absolutely crucial component of plate tectonics theory. The concept of convection in the Earth's mantle is refuted because thermal expansion cannot overcome the 62% higher density at the mantle's bottom than at its top (Figure 5).

Frequent reference is (wrongly) made to the parameters of the Earth's mantle yielding a high Rayleigh Number supposedly indicative of vigorous convection. As with the case of the Earth's core, the Rayleigh Number was derived [20] upon the basis of assumptions [21] which are inconsistent with the physical parameters of the Earth's mantle, viz.; Earth's mantle being "incompressible", density being "constant" except as modified by thermal expansion, and pressure being "unimportant" (quotes from Lord Rayleigh [20]).

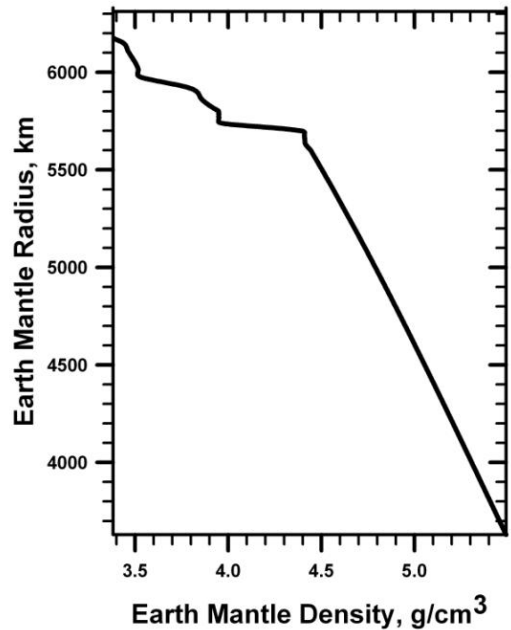

**Figure 5.** Density as a function of radius in the Earth's mantle [22].

The seeming agreement between geodynamo calculations and the observed geomagnetic field arises because, with a change of dimensions, those calculations would apply to the georeactor fluid sub-shell where convection does not violate the properties of matter. Similarly, the seeming agreement between seafloor observations and plate tectonics arises as a consequence of global dynamics as described by Herndon's Whole-Earth Decompression Dynamics [24, 25], which does not require or depend upon mantle convection.